\documentclass[structabstract]{aa}  
\usepackage{graphicx}
\usepackage{txfonts}
\usepackage{natbib}
\begin{document}
   \title{The kinematics of the diffuse ionized gas in NGC 4666}

%   \subtitle{I. No!!!}

   \author{P. Voigtl\"ander\inst{1} \and P. Kamphuis\inst{1,2}
          \and
          M. Marcelin\inst{3} \and D. J. Bomans\inst{1} \and R.-J. Dettmar\inst{1}
          }

   \institute{Astronomisches Institut, Ruhr-Universit\"at Bochum,
              Universit\"atsstr. 150, D-44801 Bochum, Germany\\
              \email{voigtlaender@astro.rub.de}
         \and
	     CSIRO Astronomy \& Space Science, P.O. Box 76, Epping, NSW 1710, Australia 
         \and
            Aix Marseille Universit\'e, CNRS, LAM (Laboratoire d'Astrophysique de Marseille) UMR 7326, 13388, Marseille, France \\}

   \date{Received 25 January 2013; accepted 11 April 2013}

% \abstract{}{}{}{}{} 
% 5 {} token are mandatory
 
  \abstract
   {The global properties of the interstellar medium with processes such as infall and outflow of gas and a large scale circulation of matter and its consequences for star formation and chemical enrichment are important for the understanding of galaxy evolution.}
   {In this paper we studied the kinematics and morphology of the diffuse ionized gas (DIG) in the disk and in the halo of the star forming spiral galaxy NGC~4666 to derive information about its kinematical properties. Especially, we searched for infalling and outflowing ionized gas.}
   {We determined surface brightness, radial velocity, and velocity dispersion of the warm ionized gas via high spectral resolution (R\,$\approx$\,9000) Fabry-P\'erot interferometry. This allows the determination of the global velocity field and the detection of local deviations from this verlocity field. We calculated models of the DIG distribution and its kinematics for comparison with the measured data. In this way we determined fundamental parameters such as the inclination and the scale height of NGC~4666, and established the need for an additional gas component to fit our observed data.}
   {We found individual areas, especially along the minor axis, with gas components reaching into the halo which we interpret as an outflowing component of the diffuse ionized gas. As the main result of our study, we were able to determine that the vertical structure of the DIG distribution in NGC 4666 is best modeled with two components of ionized gas, a thick and a thin disk with 0.8\,kpc and 0.2\,kpc scale height, respectively. Therefore, the enhanced star formation in NGC~4666 drives an outflow and also maintains a thick ionized gas layer reminiscent of the Reynold's layer in the Milky Way.}
   {}


   \maketitle
%
%________________________________________________________________

\section{Introduction}

It is well known that stellar winds and supernovae explosions are the processes that contaminate the interstellar medium (ISM) with heavy elements \citep{shields2002}. Furthermore, supernova explosions can trigger additional star formation because of the pressure they provide to the ISM. On galactic scales stellar winds and supernova explosions can influence the galactic evolution in an additional way. They locally heat and accelerate the ISM, thus creating so-called superbubbles (SBs) \citep{tomisaka1981}. Because of lower pressures in the vertical direction, these SBs expand faster in the vertical direction creating structures, much like smoking chimneys, through which the gas is able to flow into the halo of the galaxy \citep{norman1989}.\\
\indent Depending on the energy of this outflow, one can distinguish two scenarios. Either the gas has enough energy to escape the host galaxy, creating a galactic wind that enriches the intergalactic medium (IGM) \citep{heckman2002, veilleux2005}, or the gas is unable to escape the gravitational potential of the host galaxy. In the second case the chemical enrichment of the galaxy is delayed and a large gaseous halo can be created around the galaxy.\\
\indent \citet{shapiro1976} described the processes of the gas dynamics into the halo and the fall-back as a galactic fountain. In this model the ejected gas condenses to clouds in the halo and falls back onto the disk. Recently, \citet{marinacci2011} studied the galactic fountain model by using two-dimensional hydrodynamical simulations of a cloud ejected from the disk into the hot coronal gas. They described the influence of radiative cooling on the structure of this cloud and detected in their simulations a deceleration of the ejected cloud as a function of the height above the plane, i.e., a vertical velocity gradient. This is also called a lagging halo. Lagging halos are observed in several galaxies e.g., NGC~5775 \citep{heald2006} or NGC~891 \citep{fraternali2005, heald2006_2, kamphuis2007, oosterloo2007}.\\

\indent The ejected gas in the halo exists in different phases. This paper is focused on the extraplanar diffuse ionized gas (eDIG). This extraplanar component of the diffuse ionized gas (DIG) was detected in several galaxies, for example in our Milky Way \citep{Reynolds1990}. The study of the eDIG in other galaxies has been made in several galaxies in H$\alpha$ \citep{dettmar1990, Rossa2000, Rossa2003_1, Rossa2003_2, rand1990, rand1996}. Detailed kinematic studies of the halo gas were made in several dwarf galaxies \citep{martin2002, vaneymeren2009_1, vaneymeren2009_2} and continued in this paper for the SABc galaxy NGC~4666.\\
\indent Here we present a study of the kinematics of the diffuse ionized gas in the disk as well as in the halo of NGC~4666 using Fabry-P\'erot interferometry. In this context we will determine the possible consequences of the strong starburst. This includes the detection of outflows, described by \citet{dahlem1997}, and the physical properties of an eDIG layer (thick disk).\\
\indent The galaxy NGC 4666 supposedly has a high star formation rate (SFR) \citep{walter2004} of 7.0\,M$_{\sun}$\,yr$^{-1}$ \citep{dahlem1997}. It has been well studied at several wavelengths, e.g., in H$\alpha$ and R-band \citep{lehnert1995, lehnert1996_1, lehnert1996_2}, radio continuum and soft X-ray \citep{dahlem1997, dahlem1998}, and 21\,cm and CO \citep{walter2004}.\\
\indent \citet{garcia1993} determined NGC~4666 to be a member of a small group of galaxies. Besides NGC~4666 this group is composed of NGC 4632 and NGC~4668. In this context it is currently believed that the strong starburst is triggered by ongoing far-field gravitational interactions \citep{walter2004}. Table \ref{ngc4666data} shows the general parameters of the galaxy.

This paper is structured as follows: in Sect. 2 we describe the observations and the data reduction and in Sect. 3 follows the results of our analysis of the galaxy. In Sect. 4 our H$\alpha$ observations are compared to other observations, especially in other wavelengths, and in Sect. 5 we summarize our results.

   \begin{table}
      \caption[]{General parameters of NGC 4666.}
         \label{ngc4666data}
     $$ 
         \begin{array}{p{0.6\linewidth}l}
            \hline
            \noalign{\smallskip}
            Parameters (Unit)      &  \mathrm{Value} \\
            \noalign{\smallskip}
            \hline
            \noalign{\smallskip}
            optical center: 				&    \\
	    $\alpha$ (J2000.0)$^a$  			& 12^h 45^m 08.^s 6         \\
            $\delta$ (J2000.0)$^a$  			& -00^\circ 27' 43'' \\
            D [Mpc]$^a$    				& 18.225   \\
            Helioc. Radial velocity (km s$^{-1}$)$^a$		& 1529 \\
            Galaxy type$^a$				& \mathrm{SABc} \\
            Size (Maj. $\times$ Min. Diameter) (\arcmin)$^a$	& 4.6 \times 1.3 \\
	    Inclination ($^\circ$)$^b$			& 69.6\\
	    Position angle ($^\circ$)$^c$		& 42\\
	    Max. rotation velocity (km s$^{-1}$)$^b$		& 192.9 \pm 2.3\\
            \noalign{\smallskip}
            \hline
         \end{array}
     $$ 
\begin{list}{}{}
\item[$^a$] Data from NASA/IPAC Extragalactic Database (NED) (http://ned.ipac.caltech.edu/).
\item[$^b$] Data from the Lyon/Meudon Extragalactic Database (LEDA) (http://leda.univ-lyon1.fr/).
\item[$^c$] Data from \cite{springob2007}.
\end{list}
   \end{table}

\section{Observations and data reduction}

   For our observations of NGC~4666 we used the high spectral resolution (R\,$\approx$\,9000) Marseille Fabry-P\'erot interferometer (FPI) on the 1.93\,m telescope at the Observatoire de Haute-Provence (OHP) in France. The FPI was equipped with a photon counting system \citep{gach2002} which provides a field of view (FOV) of 5.4\,$\times$\,5.4\,arcmin$^2$ on 512\,$\times$\,512\,pixels$^2$. This results in a pixel size of 0.63\arcsec. Furthermore, we used a narrow band H$\alpha$ filter with a width of 10.10\,\AA~and a maximum transmission of 70\% at 6597.7\,\AA. The systemic velocity of the galaxy is 1529\,km\,s$^{-1}$ (see Table \ref{ngc4666data}) which redshifts the H$\alpha$ line to 6596.3\,\AA, close to the maximum transmission of the chosen H$\alpha$ filter. The free spectral range of the FPI is about 8.34\,\AA~or 379\,km\,s$^{-1}$. This results, using a 32-channel mode, in a sampling step of 0.26\,\AA~or 11.83\,km\,s$^{-1}$, respectively. By measuring the FWHM of a neon line, emitted by a neon calibration lamp, the finesse could be determined to 11.5 and the spectral resolution to 0.73\,\AA~or 33.0\,km\,s$^{-1}$, which is less than 3 channels.

   Because of reflections within the FPI, a ghost image of the galaxy is produced. This ghost is mirrored in the center of the FOV. To prevent an overlap of the real signal and the signal of the ghost we positioned the galaxy about 1\arcmin~offset from the center of the FOV. This offset is small enough that the galaxy does not leave the FOV, but is large enough to ensure a clear separation between the ghost and the galaxy, thus preventing any contamination.

   The observations were obtained in two nights in early February 2011. In these two nights the galaxy was observed in 50 and 47 cycles, respectively. A failure in the cooling system increased noise and created bad cycles, which were then removed, leaving us with 50 and 44 cycles for each night. To identify possible extended halo structures we moved the telescope for the second observation about 1.7\arcmin~west in order to position the ghost image once east and once west of the galaxy. With an integration time of 10\,s per channel in one cycle we reached a total integration of 30,080\,s ($\approx$ 8.36\,h) or 940\,s ($\approx$ 15.7\,min.) per channel.
  
   The data were reduced with the {\tt IDL} program {\tt FPreduc} \citep{daigle2006, epinat2009}. Both cubes were reduced independently. In order to correct for fluctuations in the telescope positioning, the position of every 10~s exposure was corrected by matching the positions of several stars in the FOV. Furthermore, the sky lines were subtracted and a heliocentric velocity correction was applied. With the {\tt koords} routine of the {\tt karma} package \citep{gooch1996} both cubes were provided with a coordinate system by matching four stars in the FOV to their counterparts in a Digitized Sky Survey (DSS) image.

After the reduction with {\tt FPreduc} the continuum was still visible in the data. Therefore,  the continuum was subtracted using the continuum image which was produced during the data reduction. Once both cubes were fully reduced they were moved to the same pixel coordinates and combined.

The observed velocities of NGC 4666 cover a larger range than the available free spectral range. Therefore, some areas with extreme velocities appear in channels that are offset from their true velocities by the spectral range. These velocities are measured from the previous or next order of the FPI. In order to correct this, the cube was extended with 16 channels, 10 channels in front and 6 channels behind the normal velocity range. Such parts of emission which contained velocities of the previous order were added in the channels behind the normal range and such parts of emission which contained velocities of the next order were added in front of the normal range. Furthermore, these regions were blanked within the normal range.

The subsequent analysis of the data cube was done with the {\tt GIPSY} package \citep{vanderhulst1992}. Although {\tt GIPSY} is mainly used for the analysis of interferometric observations of \ion{H}{i} line emission, it has many functions that can also be applied to FPI observations of H$\alpha$ emission. To optimize the S/N the cube was spatially smoothed by using the task {\tt SMOOTH}, from 3.6\arcsec, which is equal to the average seeing, to a size of 5.0\arcsec. This value of 5.0\arcsec~was determined to be the best compromise between increasing the S/N in the outer parts of the galaxy and a good spatial resolution. For the following we define the number of complete pixels within a block of 5.0\arcsec\,$\times$\,5.0\arcsec as one resolution element, which corresponds to 7\,$\times$\,7 pixels.

%__________________________________________________________________

\section{Analysis}
In our analysis we focus on five different areas. Initially, we determine the sensitivity of our observation, then we verify the statement of \citep{walter2004} that NGC~4666 is a starburst galaxy by determining the star formation rate. The third step is the velocity analysis of our data. From this analysis we obtain a rotation curve for NGC~4666 and a good impression of the kinematics. In order to understand how much of the gas is in the disk and how much is in the halo, we use this kinematical information to construct a model for the disk. Finally, we compare our disk model to the data to understand structures in the halo of the galaxy.

\subsection{Sensitivity of our observation and verification of the star formation rate}

To determine the sensitivity of our data we added all channels to generate an integrated H$\alpha$ map which we will refer to as the intensity map from here on. With a total H$\alpha$ flux of 4.7\,$\times$\,$10^{-12}$\,erg\,s$^{-1}$\,cm$^{-2}$ of NGC~4666, determined by \citet{moustakas2006} with drift-scanning spectroscopy, we determined the 3$\sigma$ sensitivity of our observations to be 5.6\,$\times$\,$10^{-18}$\,erg\,s$^{-1}$\,cm$^{-2}$\,arcsec$^{-2}$. Another common unit to describe the intensity of emission is the so-called emission measure (EM). It is given by

\begin{equation}
 \textrm{EM} = \int{n_e^2} dr,
 \label{eq: em}
\end{equation}
where n$_e$ is the electron density and r the length along the line-of-sight. For a direct calculation we could use the formula \citep{dettmar1992_2}

\begin{equation}
 \textrm{EM} = 5\,\times\,10^{17}\,F_{H\alpha}/\Omega\,\textrm{cm}^{-6}\,\textrm{pc},
 \label{eq: em2}
\end{equation}
with the measured flux F$_{H\alpha}$ in erg\,cm$^{-2}$\,sec$^{-1}$ and $\Omega$ the collecting area in $\square\arcsec$. This results in an emission measure of EM~=~2.8\,cm$^{-6}$\,pc for the aforementioned sensitivity.

NGC~4666 is described as a superwind galaxy with a high SFR of 7.0\,M$_{\sun}$\,yr$^{-1}$ \citep{dahlem1997} or even as a starburst galaxy \citep{walter2004}. This value should also be verified for our discussion. For H$\alpha$ we took the aforementioned flux of 4.7\,$\times$\,$10^{-12}$\,erg\,s$^{-1}$\,cm$^{-2}$ and for comparison the flux density in FIR at 60\,$\mu$m of 37.34 Jy \citep{soifer1989}. The H$\alpha$ flux is influenced by extinction. To obtain the extinction, we used the H$\beta$ flux of 0.90054\,$\times$\,$10^{-12}$\,erg\,s$^{-1}$\,cm$^{-2}$ determined by \citet{kennicutt2009}, and calculated a Balmer decrement of $\frac{H\alpha}{H\beta}~=~5.22$ which is above the expected ratio of 2.86 for a temperature of 10,000\,K \citep{osterbrock1989}. Taking into account the reddening law \citep{fitzpatrick1999}, the formula for the color excess E(B-V) is given by \citet{xiao2012} as

\begin{equation}
 \textrm{E(B-V)} = 1.99 \times \log \frac{H\alpha/H\beta}{2.86}.
\end{equation}

\noindent This results in a color excess of E(B-V)~=~0.52 which corresponds to A(V)~=~1.61, with A(V)~=~3.1\,$\times$\,E(B-V), and A(B)~=~2.13. With A(H$\alpha$)~=~0.538\,$\times$\,A(B) the extinction of the H$\alpha$ line can be calculated as A(H$\alpha$)~=~1.15. Therefore, the total H$\alpha$ flux is increased by a factor of 2.88 to 13.51\,$\times$\,$10^{-12}$\,erg\,s$^{-1}$\,cm$^{-2}$. For the additional calculations, we used the mean distance of 18.2\,Mpc, provided in NED\footnote{NASA Extragalactic Database, {\tt http://ned.ipac.caltech.edu/}}, whereas \citet{dahlem1997} used 26.3\,Mpc. The different values given by NED are mostly obtained by the Tully-Fisher relation and are approximately 19\,Mpc. The final value of NED we used is the average of all these single values. In contrast, the value of \citet{dahlem1997} is based on the redshift with a correction for the Virgo infall. Since both methods are not very accurate we prefer to take the average from NED. Using the formulas of \citet{kennicutt1998}, we obtained an SFR of
\begin{equation}
 \textrm{SFR} = 7.9 \times 10^{-42} L_{H\alpha} \frac{\textrm{M}_{\sun}\,\textrm{yr}^{-1}}{\textrm{erg\,s}^{-1}} = 4.24\,\textrm{M}_{\sun}\,\textrm{yr}^{-1}
\end{equation}
for H$\alpha$ and 
\begin{equation}
 \textrm{SFR} = 4.5 \times 10^{-44} L_\textrm{FIR} \frac{\textrm{M}_{\sun}\,\textrm{yr}^{-1}}{\textrm{erg\,s}^{-1}} = 2.2\,\textrm{M}_{\sun}\,\textrm{yr}^{-1}
\end{equation}
for far-infrared (FIR).

Using the distance of 26.3\,Mpc \citep{dahlem1997} the SFRs would increase by a factor of about 2. This results in a good agreement between our H$\alpha$ SFR and value of 7.0\,M$_{\sun}$\,yr$^{-1}$ determined by \citet{dahlem1997}. The remaining differences between the SFR in H$\alpha$ and FIR can easily be explained by galaxy specific characteristics \citep{kennicutt1998}. In this context \citet{kennicutt1998} referenced \citet{buat1996} who determined a coefficient of 8$^{+8}_{-3}\,\times\,10^{-44}$ instead of the above-mentioned 4.5\,$\times\,10^{-44}$.

\subsection{Kinematic analysis}
\begin{figure*}
   \centering
   \includegraphics[angle=0, width=0.98\textwidth]{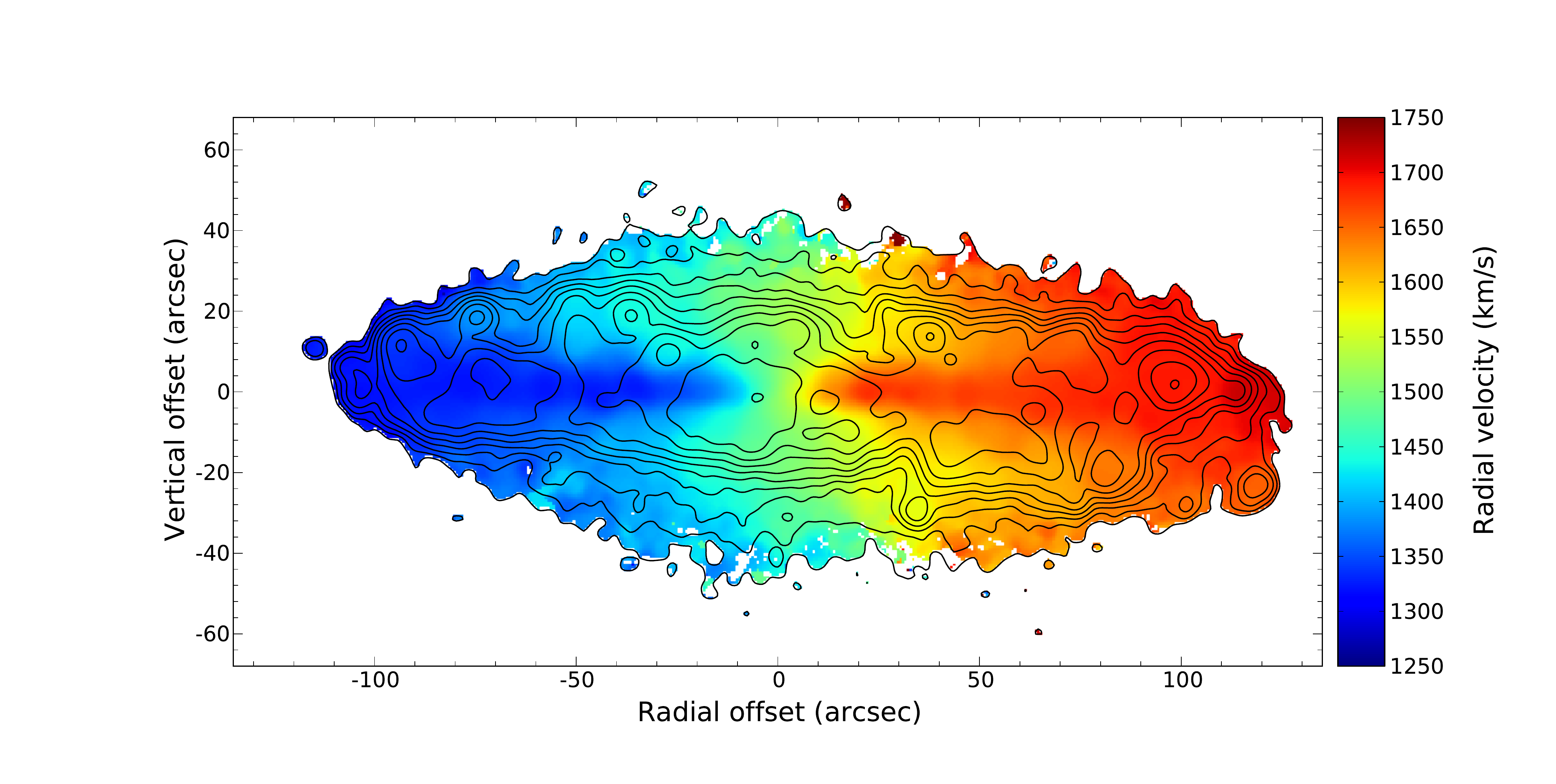}
      \caption{Velocity map of NGC~4666 with coordinates in offsets from the center of the galaxy in arcseconds. The contours show the intensity map at 3$\sigma$, 6$\sigma$, 12$\sigma$, 24$\sigma$, etc.}
         \label{ngc4666velocity}
   \end{figure*}

\begin{figure*}
   \centering
   \includegraphics[angle=0, width=0.98\textwidth]{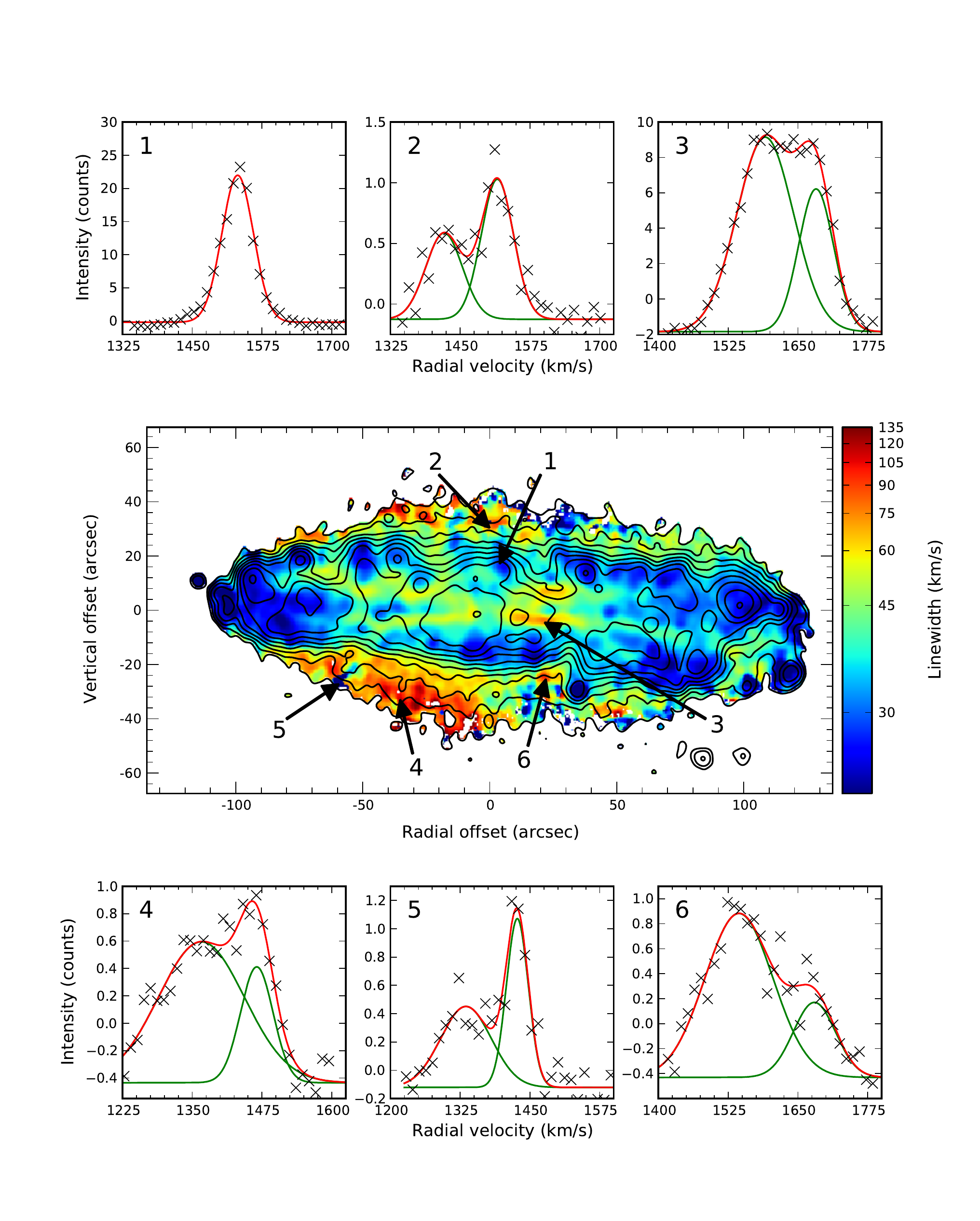}
      \caption{The linewidth map of NGC~4666, resulting from a single Gaussian fit, with coordinates given in distances in arcseconds to the center of the galaxy. The contours show the intensity map at 3$\sigma$, 6$\sigma$, 12$\sigma$, 24$\sigma$, etc. Furthermore, some profiles of chosen pixels are displayed to show that high linewidths are double Gaussian profiles in the selected areas. When there are two Gaussians, the single Gaussian profiles are displayed in green and the sum of both in red.}
         \label{ngc4666dispersion}
   \end{figure*}

Before starting the further analysis of the data cube the above-mentioned ghost images of the galaxy below and above the disk were removed. For the further analysis the {\tt GIPSY} task {\tt XGAUFIT} was used. First a single Gaussian fit of every pixel with {\tt XGAUFIT} was performed by using blocking filters for velocity and dispersion/linewidth\footnote{We define the linewidth as the dispersion of the Gaussian fit of the data which is not corrected for the instrumental dispersion.}. For the velocity only values within the velocity range of the extended cube were accepted. For the linewidth the minimum value was set to the instrumental dispersion of 14\,km\,s$^{-1}$, which follows directly from the above-mentioned spectral resolution of the instrument. The maximal linewidth was set to 162\,km\,s$^{-1}$, which is equal to the dispersion of a Gaussian with a FWHM of the free spectral range. All these filters suppress unphysical fitting results. The intensity map was masked at the 3$\sigma$ level and subsequently used as a mask for the velocity field. As a last processing step a filter that blanked any isolated pixels was applied to the data. This results in a well-cleaned velocity field which can be seen in Fig.~\ref{ngc4666velocity}. All these steps were also done for the linewidth map which was produced with {\tt XGAUFIT}, resulting in Fig.~\ref{ngc4666dispersion} \textit{(center)}. In both figures the black contours describe the intensity at 3$\sigma$, 6$\sigma$, 12$\sigma$, 24$\sigma$, etc.

Because of our predominant interest in the vertical structure, the velocity map and the linewidth map were rotated so that the major axis of the galaxy is parallel to the x-axis. Moreover, the global coordinate system in right ascension and declination was replaced by a relative coordinate system that describes the radial and vertical offset to the center of NGC~4666.

These maps of NGC~4666 show a lopsidedness with a radial extent of the approaching part of about 110\arcsec~and a radial extent of the receding part of about 125\arcsec.
Furthermore, the velocity map immediately shows several features beyond this lopsidedness. When comparing the intensity contours in the velocity map, one can see the influence of the \ion{H}{ii} regions to the measured velocities, e.g., at the \ion{H}{ii} regions at a radial/vertical offset of -74\arcsec/16\arcsec~and -30\arcsec/8\arcsec.

\begin{figure}
   \centering
   \includegraphics[width=0.49\textwidth]{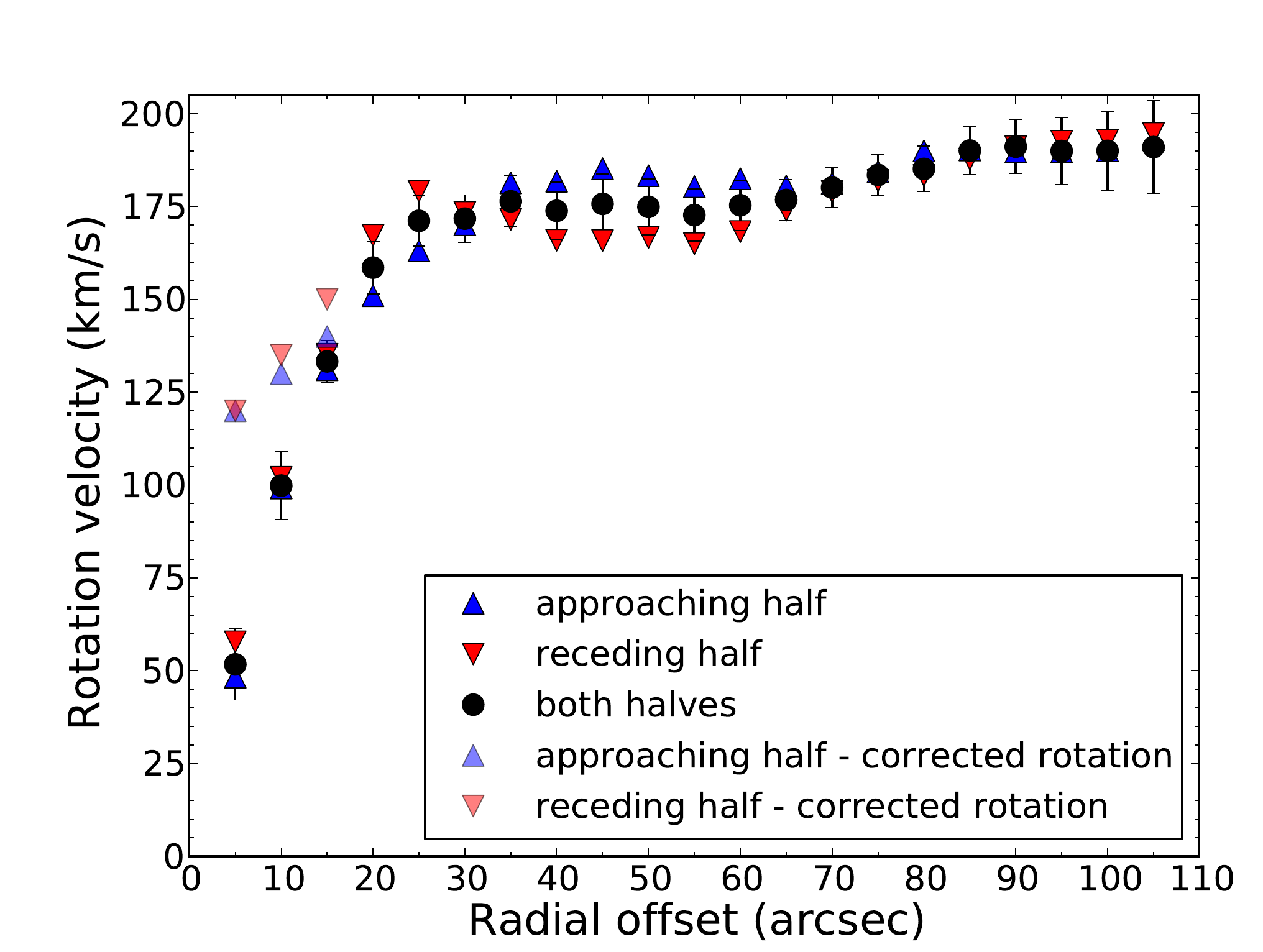}
      \caption{The rotation velocity vs. the radial distance from the center. The blue triangles describe the approaching half of the galaxy, the red triangles the receding half, and the black dots both halves in a common calculation with the sigma of the velocity given by {\tt ROTCUR} as error bars. The semi-transparent triangles in blue and red describe the corrected rotation of the approaching and the receding half for the final model, respectively.}
      \label{ngc4666vrot}
   \end{figure}

\begin{figure}
   \centering
   \includegraphics[width=0.49\textwidth]{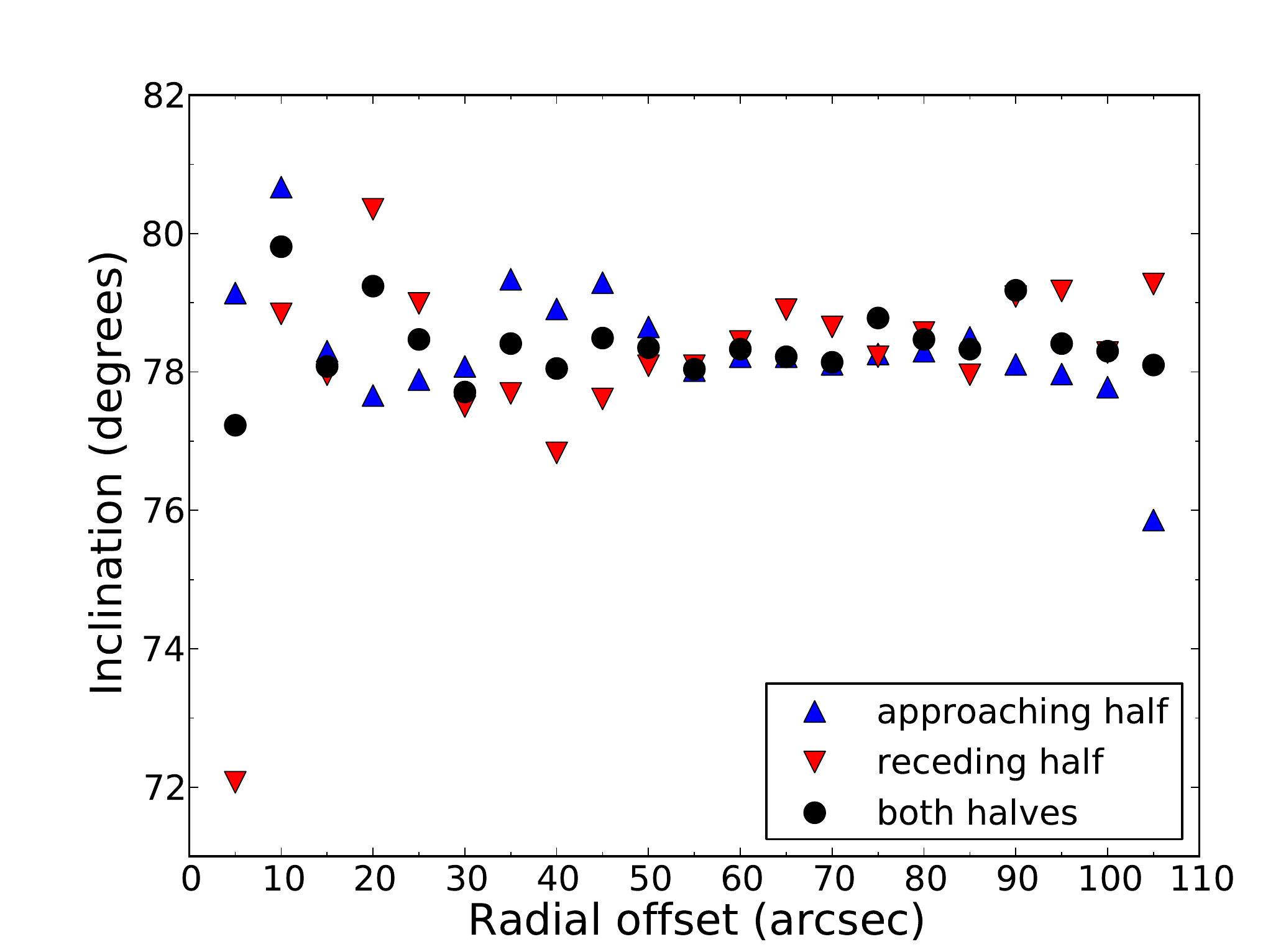}
      \caption{The inclination vs. the radial distance from the center. Symbols as in Fig. \ref{ngc4666vrot}.}
      \label{ngc4666incl}
   \end{figure}

To determine the position angle and kinematical center and to obtain a rotation curve for NGC~4666, a tilted ring model was fitted by means of $\chi^2$ minimization with the {\tt GIPSY} task {\tt ROTCUR} on the velocity map. Because of the lopsidedness of NGC~4666, only the inner 105\arcsec~could be fitted well. The width of the rings was set to 5\arcsec, our spatial resolution after smoothing. In a first run the position of the center of the galaxy, the rotation velocity, the systemic velocity, the inclination, and the position angle were all fitted in order to obtain initial estimates. Step-by-step single parameters were fixed using the average value over all the rings to improve the fit for the other parameters. In the last iteration only the rotation velocity and the inclination were not fixed. The results are shown in Figs.~\ref{ngc4666vrot} and \ref{ngc4666incl} for the rotation curve and the inclination, respectively. All the other parameters and the average of the inclination are given in Table \ref{ngc4666rotcur}.

   \begin{table}
      \caption[]{Results of {\tt ROTCUR} for NGC 4666. The values are the averages of the inner $\pm$105\arcsec.}
         \label{ngc4666rotcur}
     $$ 
         \begin{array}{p{0.6\linewidth}l}
            \hline
            \noalign{\smallskip}
            Parameters (Unit)      &  \mathrm{Value} \\
            \noalign{\smallskip}
            \hline
            \noalign{\smallskip}
            kinematical center: 				&    \\
	    $\alpha$ (J2000.0)  			& 12^h 45^m 08.^s 6.8         \\
            $\delta$ (J2000.0)  			& -00^\circ 27' 43.9'' \\
            Systemic velocity (km s$^{-1}$)	& 1506\pm 4\\
	    Inclination ($^\circ$)		& 78.3 \pm 0.5\\
	    Position angle ($^\circ$)		& 222.6 \pm 0.8\\
            \noalign{\smallskip}
            \hline
         \end{array}
     $$ 
   \end{table}

Looking at Fig. \ref{ngc4666vrot} one can see an almost flat rotation curve with a maximal rotation velocity of about 190-195\,km\,s$^{-1}$ which agrees well with the value found in the literature given in Table~\ref{ngc4666data}. The position of the galactic center and the position angle also correspond to the values found in the literature (see Table \ref{ngc4666data}). The value of the position angle shows this offset of about 180$^\circ$ because of the definition in {\tt ROTCUR}; in {\tt ROTCUR} the position angle is defined as the counterclockwise angle between the north direction and the major axis of the receding part of the galaxy. The systemic velocity, the inclination and the position angle are the average values over all the rings up to 105\arcsec~with the standard deviation as the error.

\begin{figure}
   \centering
   \includegraphics[width=0.49\textwidth]{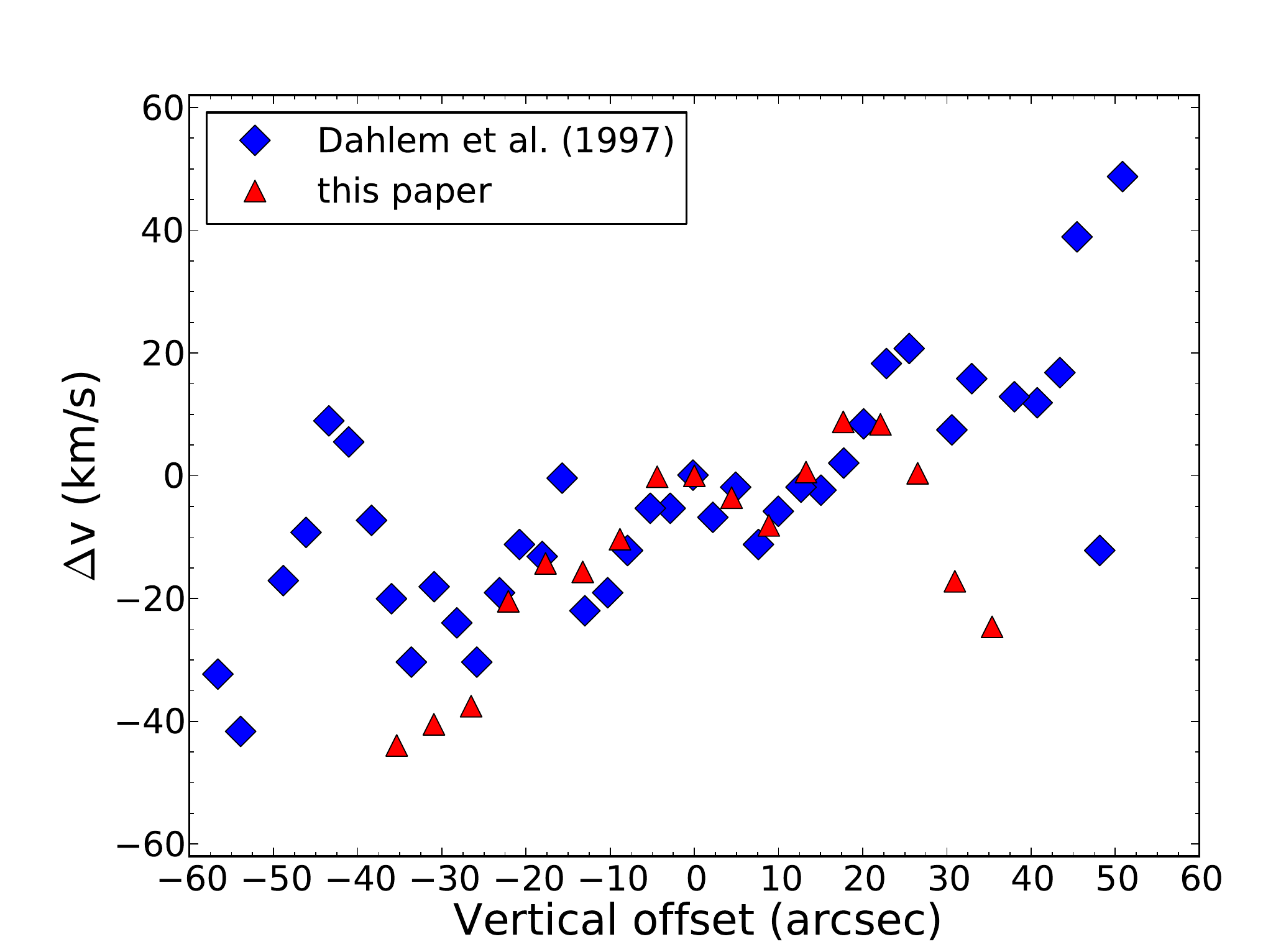}
      \caption{The velocity offset along the minor axis of NGC~4666. The red triangles show our data; the blue diamonds show the data of \citet{dahlem1997}. For this comparison the sign of the vertical offset in the data of \citet{dahlem1997} was switched because of the different convention of the position angle.}
      \label{ngc4666velminaxis}
   \end{figure}

In the next step, the velocity offset along the minor axis of up to 100 km s$^{-1}$ described by \citet{dahlem1997} will be investigated. To verify this in our data we determined the velocity for every resolution element along the minor axis. Thus, we obtained for every vertical offset y along the minor axis a $\Delta$v which is defined as $\Delta$v = v$_y$ - v$_0$, with v$_y$ as the velocity at a given vertical offset and v$_0$ as the velocity of the central resolution element (1510\,km s$^{-1}$). In Fig. \ref{ngc4666velminaxis} we plot $\Delta$v vs. the vertical offset of our data in comparison to the data of \citet{dahlem1997}. To do this, we switched the sign of the vertical offset in the data of \citet{dahlem1997} because of the 180$^\circ$ difference in the assumed PA.

In direct comparison, the general shape of the velocities between -20\arcsec~and +20\arcsec~vertical offset agrees well with the one presented in \citet{dahlem1997}. The velocity variations in the \citet{dahlem1997} data and in our data in this inner area can be explained by the influence of \ion{H}{ii} regions. The differences in the outer part are probably a combined effect of reduced S/N and a difference in velocity resolution ($\lambda_{\textrm{res}}$~=~0.73\,\AA~$\cor$~33\,km\,s$^{-1}$ in this work, $\lambda_{\textrm{res}}$~=~4\,\AA~$\cor$~182\,km\,s$^{-1}$ in \citet{dahlem1997}).

To study further this velocity offset in the outer regions and to detect other anomalies in the kinematics of the galaxy, the linewidth of the Gaussian fit is a useful tool. Therefore, we have chosen six areas to describe features in this linewidth map in more detail. The profiles of these regions can be seen in Fig.~\ref{ngc4666dispersion} around the linewidth map. First, the general linewidth is about 30-35 km s$^{-1}$, shown by way of example at position 1. The profile can be well fitted by one Gaussian (red line). Where we have compact \ion{H}{ii} regions (black contour lines) the dispersion is about 30\,km s$^{-1}$ or lower. In this linewidth map we see a region of large linewidth at a vertical offset of about 30\arcsec~along the minor axis (position 2). The profile at this position shows a double Gaussian structure with a smaller Gaussian with a central position at 1421\,km\,s$^{-1}$ and a bigger one at 1517\,km\,s$^{-1}$ (green lines), which is in good agreement with our central velocity, and also in better agreement with \citet{dahlem1997} for this vertical offset (see Fig. \ref{ngc4666velminaxis}). By fitting merely one Gaussian to the line profile our central position is shifted to a lower value due to this first component at 1421\,km\,s$^{-1}$. Without that component the velocity keeps roughly constant for a positive vertical offset. This would be expected for a normal flat rotating galaxy. The first component with a velocity of about 100\,km\,s$^{-1}$ below the systemic velocity can be interpreted as an outflowing component of the gas.

There are several other conspicuous areas in the linewidth map. In the center there is an x-shaped region with a larger linewidth which also shows a double Gaussian structure (e.g., at position 3). The linewidth map shows a linewidth for this structure up to 60-80 km s$^{-1}$. Furthermore, there is a significant rise of the linewidth, up to 50\,km\,s$^{-1}$, in the outer part of the galaxy above 20\arcsec~and below -20\arcsec~ vertical offset. In the half of the galaxy with a negative radial offset the linewidth rises even up to 130\,km\,s$^{-1}$, e.g., at position 4. Within these regions of large linewidths is a small hook-like structure where the linewidth drops down to 30-40\,km\,s$^{-1}$, marked as position 5. A detailed view of this position shows a similar shape as position 4, but here the amplitude of the first component, at 1335\,km\,s$^{-1}$, is significantly lower than the second component at 1427\,km\,s$^{-1}$ and close to the noise level. Therefore, the fit was dominated by the second component which has a small linewidth. Moreover, there is a region similar to position 2 where the linewidth rises within a compact area, marked as position 6 in Fig. \ref{ngc4666dispersion}. Both areas, 5 and 6, will be discussed in more detail later in comparison to another independent H$\alpha$ observation. 

The negative baselines of some profiles are a result of over-subtracted sky lines and because the continuum subtraction was optimized for the galactic center. Because of the over-subtraction, the above-mentioned calculated sensitivity is just an upper limit, but it does not affect the further profile analysis because it is merely a shift along the intensity axis and does not influence the shape of the profile. To understand the origin of all these double Gaussians we constructed a set of models to compare with the data. We will describe these models in the following section.\\

\subsection{Models}
To explain the aforementioned features in the velocity map and especially in the linewidth map and to determine the corrected rotation curve, an accurate model of NGC~4666 is essential. To model the data cube we used the {\tt FORTRAN} code described in \citet{kamphuis2007}. In this code the emission is calculated for every position in an exponential disk while taking the line of sight velocities into account. 

The model takes the following characteristic parameters as input. The intrinsic shape of the ionized gaseous disk is characteristed by an exponential scale length and scale height, as well as a truncation radius. The velocity structure is determined from an input rotation curve and the given linewidth. Finally, the disk is orientated according to a given inclination.

The following parameters were determined directly from the data. The linewidth was measured from the diffuse gas in Fig.~\ref{ngc4666dispersion} as 35\,km\,s$^{-1}$. The initial rotation curve and inclination were taken from the output of {\tt ROTCUR}. The truncation radius could be determined easily from the total flux map as 130\arcsec, which corresponds to 11.5\,kpc assuming a distance of 18.2 Mpc (see Table 1).

For the scale length and the scale height we started with some intitial estimates tracing the exponential intensity decrease along the major and minor axis, respectively. However, because of the large number of \ion{H}{ii} regions in the disk of NGC~4666, these values are very uncertain and had to be determined more accurately through a visual comparison of the data and the models.

As a first step in fitting the model to the data, we corrected the rotation curve and determined an accurate value for the scale length by comparing the PV-diagrams along the major axis of the data cube and the model cube to each other iteratively. This resulted in an exponential scale length of 3.0\,kpc and in corrected rotation velocities for the inner 15\arcsec~(see Fig.~\ref{ngc4666vrot}). This correction is not unexpected as we did not perform an envelope tracing \citep{sofue2001} to determine the initial rotation curve.

\begin{figure*}
   \centering
   \includegraphics[angle=90, width=0.98\textwidth]{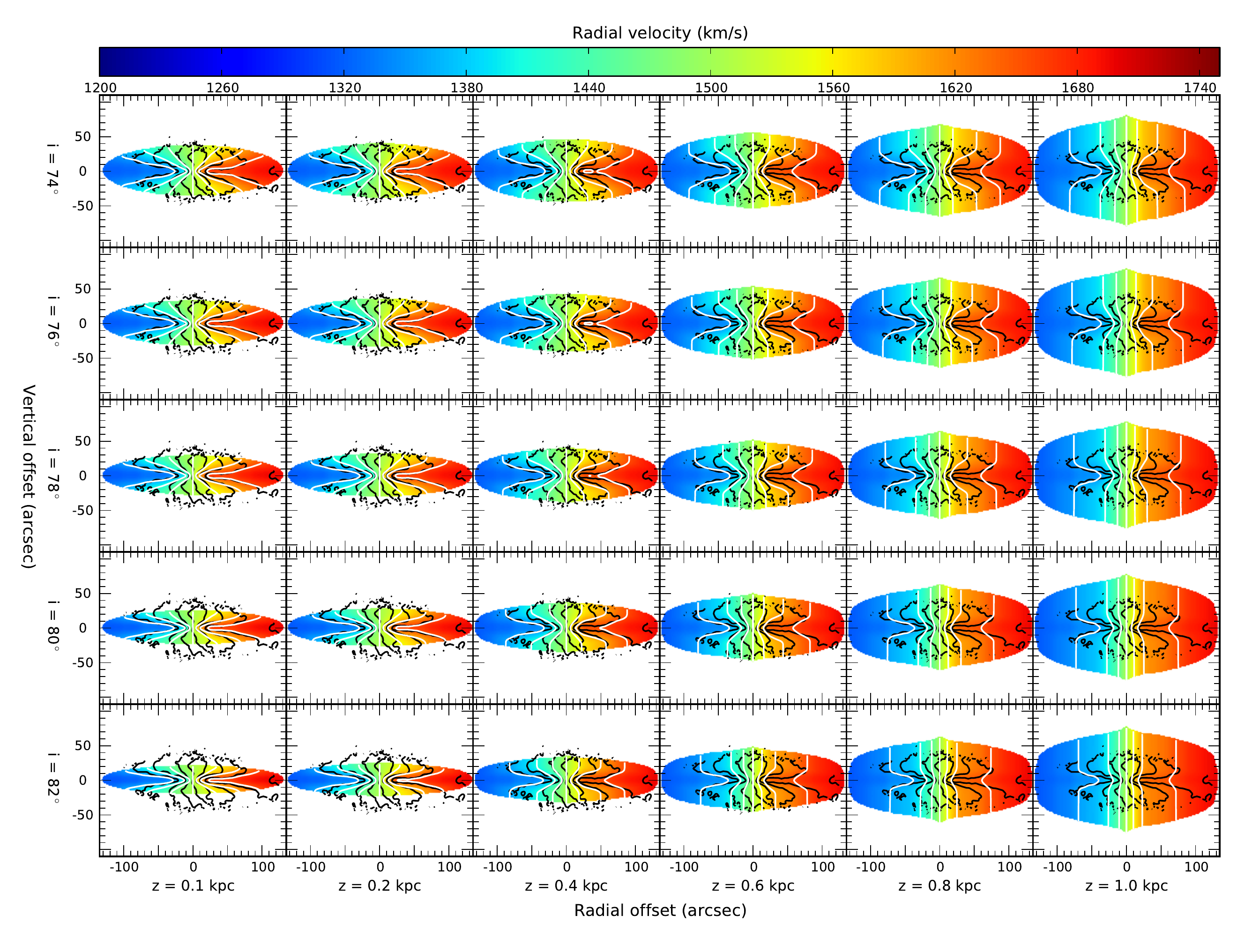}
      \caption{Velocity maps of models with white contours in steps of 50\,km s$^{-1}$ around the systemic velocity of 1506\,km s$^{-1}$ at different inclination (y-axis) and scale heights (x-axis). The black contours describe the data with the same velocity steps.}
         \label{model_velocity}
   \end{figure*}

\begin{figure*}
   \centering
   \includegraphics[angle=90, width=0.98\textwidth]{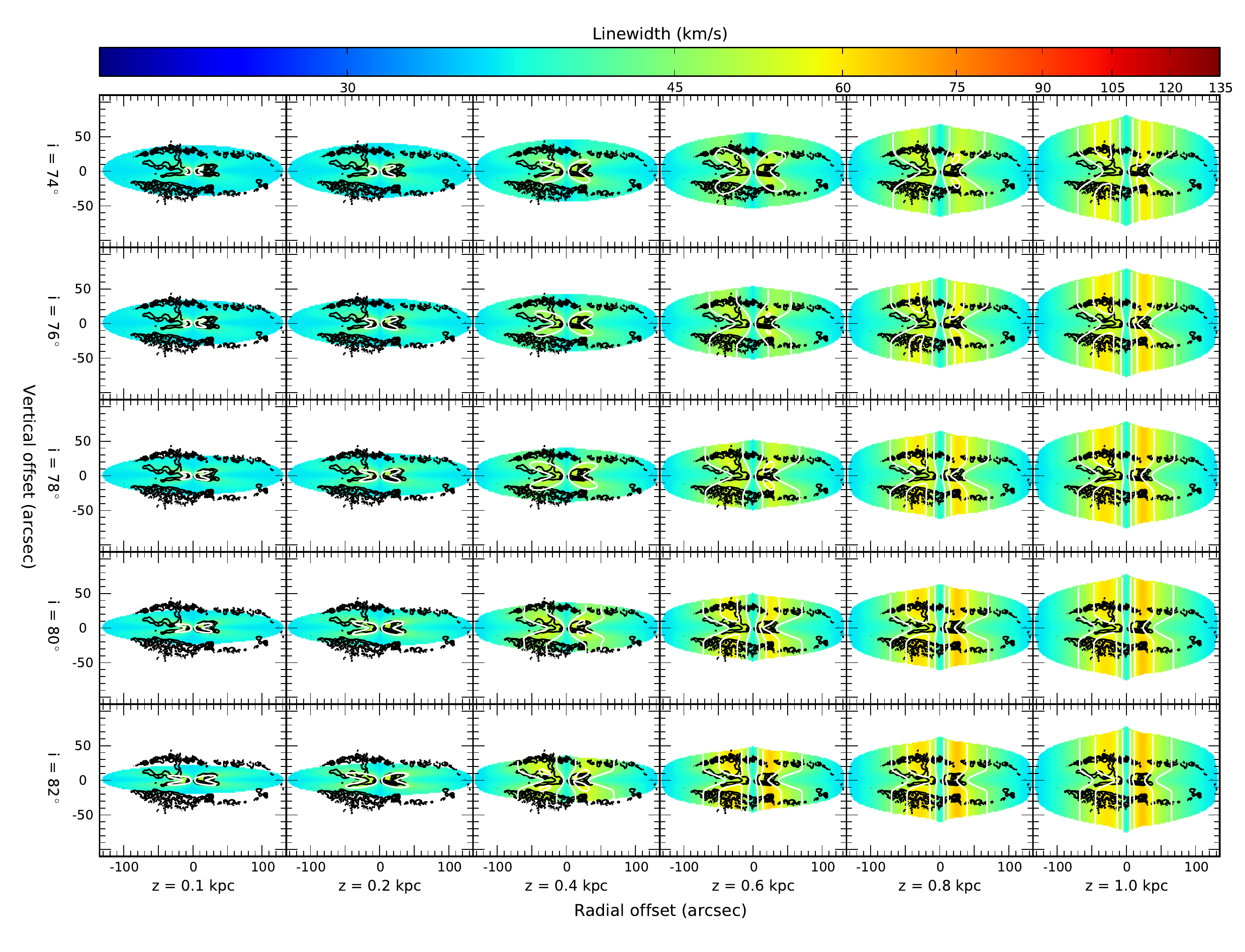}
      \caption{Linewidth maps of models with white contours at 45\,km s$^{-1}$, 55\,km s$^{-1}$, 65\,km s$^{-1}$, and 75\,km s$^{-1}$ at different inclination (y-axis) and scale heights (x-axis). The black contours describe the data with the same linewidth steps.}
         \label{model_dispersion}
   \end{figure*}

Subsequently, we determined accurate values for the inclination and the exponential scale height by calculating a grid of different models with different values for both parameters. This grid ranges from a scale height of 0.1\,kpc to 1.0\,kpc and from an inclination of 74$^\circ$ to 82$^\circ$. The resulting model cubes were analyzed in the same way as the data cube which resulted in a sample of radial velocity and linewidth maps, as seen in Figs.~\ref{model_velocity} and \ref{model_dispersion}, respectively.

\begin{figure*}
   \centering
   \includegraphics[angle=0, width=0.98\textwidth]{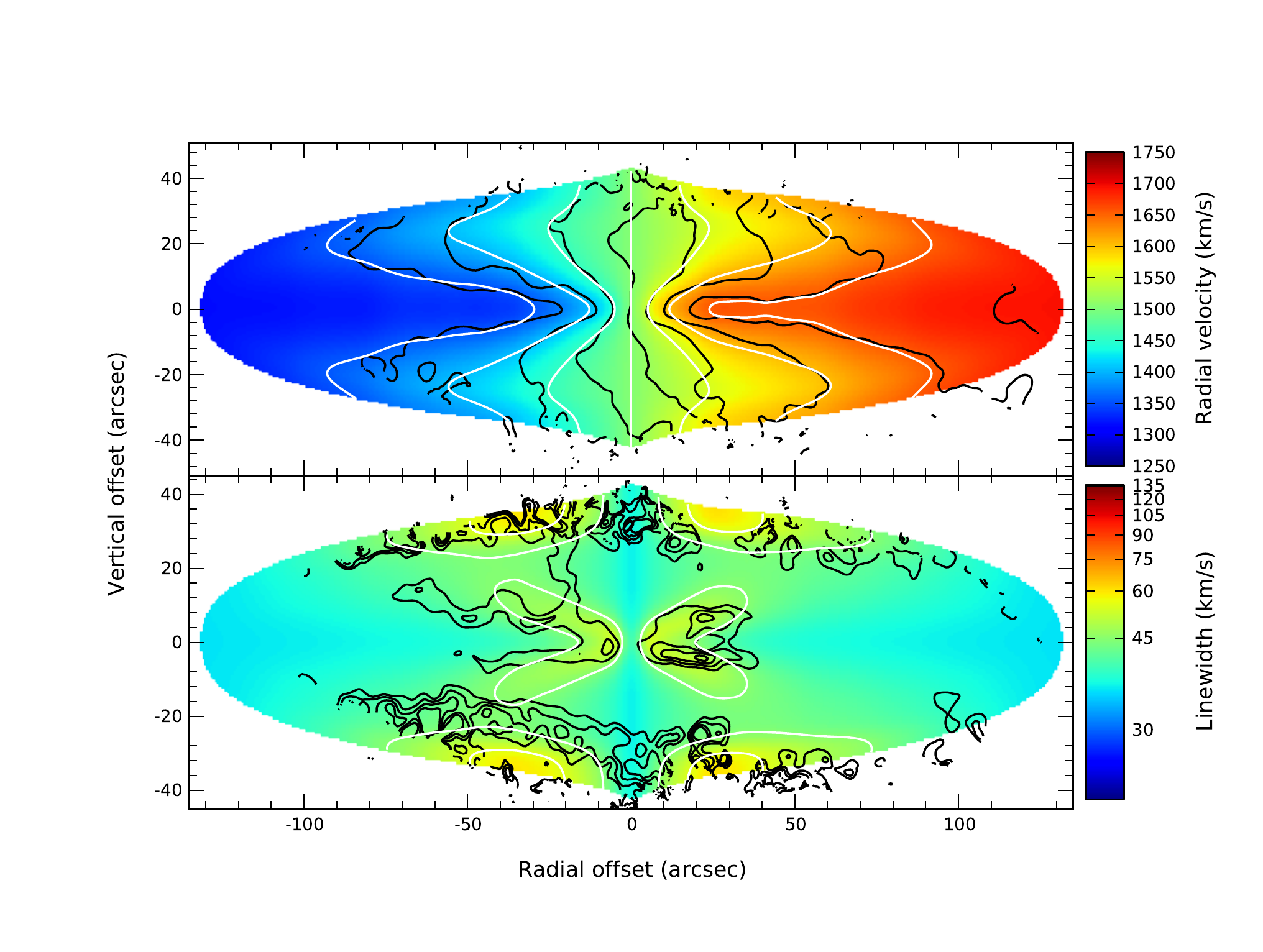}
      \caption{The final velocity model (top) and the final linewidth model (bottom). The white contours show the model and the black contours the data. In the velocity map the contours are in steps of 50\,km s$^{-1}$ around the systemic velocity of 1506\,km s$^{-1}$. In the linewidth map the contours are at 45\,km s$^{-1}$, 55\,km s$^{-1}$, 65\,km s$^{-1}$, and 75\,km s$^{-1}$.}
         \label{ngc4666veldisp_final}
   \end{figure*}

We compared the models to the data through visual inspection of velocity fields and linewidth maps. This confirmed an inclination of 78$^\circ$ on the one hand and revealed a more complex structure of the scale height on the other hand. The inner part of the galaxy matched well with the model with a scale height of 0.2\,kpc, whereas the outer part matched well with the model with a scale height of 0.8\,kpc. Therefore, we tried a combination of these two models adding the thicker disk (z~=~0.8\,kpc) to the thin disk (z~=~0.2\,kpc). We found that a flux ratio between the thin disk and the thick disk of 2.5 matched the data best. A comparison of this final model with the data is shown in Fig.~\ref{ngc4666veldisp_final} for the velocity and linewidth map and confirms the quality of the calculated model. The parameters used in the final model are summarized in Table~\ref{ngc4666model}.

We also investigated models with a dust layer and a lagging halo to study the possible influences of such components. However, these models did not show any indication of the presence of a strong dust layer or a lagging halo.

   \begin{table}
      \caption[]{Parameters for the best-fit model.}
         \label{ngc4666model}
     $$ 
         \begin{array}{p{0.6\linewidth}l}
            \hline
            \noalign{\smallskip}
            Parameters (Unit)      &  \mathrm{Value} \\
            \noalign{\smallskip}
            \hline
            \noalign{\smallskip}
            Exponential scale length (kpc)		& 3.0 \\
            Exponential scale height (kpc)		& 0.2/0.8 \\
            Inclination	($^\circ$)& 78 \\
	    Truncation radius (kpc)& 11.5\\
	    Distance (Mpc)				& 18.2\\
	    General linewidth (km\,s$^{-1}$)			& 35\\
            \noalign{\smallskip}
            \hline
         \end{array}
     $$ 
   \end{table}

In Fig.~\ref{ngc4666veldisp_final} the white contours describe the model and the black contours the data. In the velocity map (Fig.~\ref{ngc4666veldisp_final}, top) one can see again offsets in velocity along the minor axis in the data. Otherwise, there is good agreement between model and data. All the residuals can be explained by the influence of \ion{H}{ii} regions except the upper-right quadrant, especially at a vertical offset of 20\arcsec. The velocity in the data there is about 50\,km\,s$^{-1}$ higher than expected from the model. In comparison with our linewidth map (Fig.~\ref{ngc4666dispersion}), one can see a small rise of the linewidth; however, it remains in good agreement with our model. A possible explanation of this velocity offset could be a warp.

In the linewidth map of the final model (Fig. \ref{ngc4666veldisp_final}, bottom) one can see an x-shaped structure in the central region with a maximum linewidth of 45\,km\,s$^{-1}$ whereas the linewidths in the outer parts show values of up to 60\,km\,s$^{-1}$. Compared to our data, the general structure is in good agreement. Nevertheless, there are two important differences. First, in our data the linewidth in the center rises to 75\,km\,s$^{-1}$ and is therefore 30\,km\,s$^{-1}$ higher than in the model. Second, the rising linewidth in the outer part is well described by the model for a positive radial offset; for a negative radial offset the linewidth rises to 120\,km\,s$^{-1}$, which is not described by the model. Increasing the inclination of the thin disk or thick disk does not solve this problem. We conclude that we are able to reproduce an increasing linewidth in an x-shaped structure in the center of the galaxy, but we are not able to reach the same strength. 

\begin{figure}
   \centering
   \includegraphics[angle=0, width=0.49\textwidth]{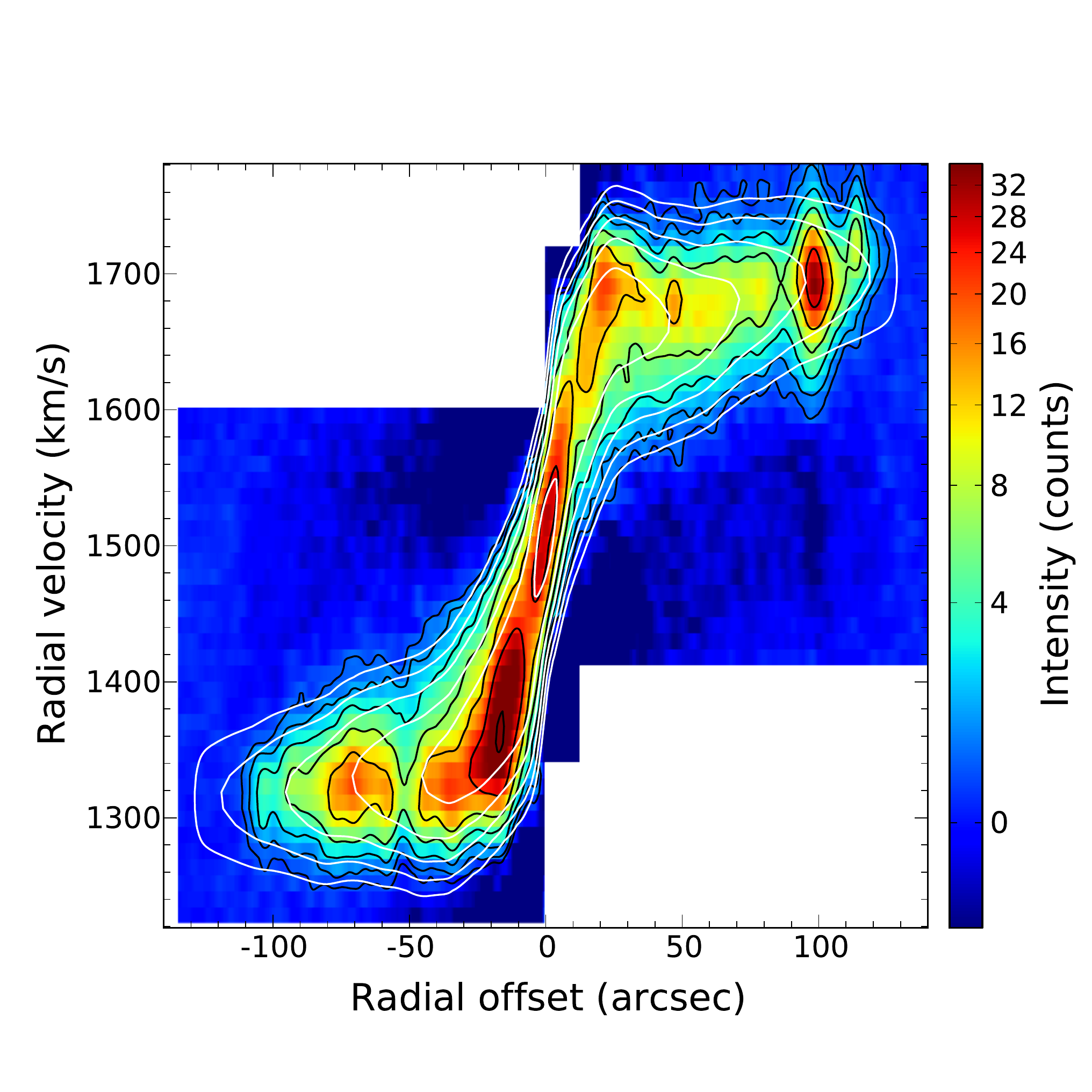}
      \caption{The PV-diagram of NGC 4666 along the major axis with the data in black and the final model in white contours.}
         \label{pv_major_axis}
   \end{figure}

To study the quality of our model Fig. \ref{pv_major_axis} shows the PV-diagram along the major axis. The chosen scale length of 3.0\,kpc agrees well with the data, but the assumed dispersion of 35\,km\,s$^{-1}$ and the flat rotation velocity of 195\,km\,s$^{-1}$ also agrees well.

As a result we conclude that the regions of large linewidth in our data, marked by way of example with 3 and 4 in Fig.~\ref{ngc4666dispersion}, are partly explained as projection effects of a thick and a thin disk in our data. For a complete explanation a detailed analysis and a comparison with other observations are necessary.

\subsection{Halo features}
\begin{figure*}
   \centering
   \includegraphics[angle=0, width=0.98\textwidth]{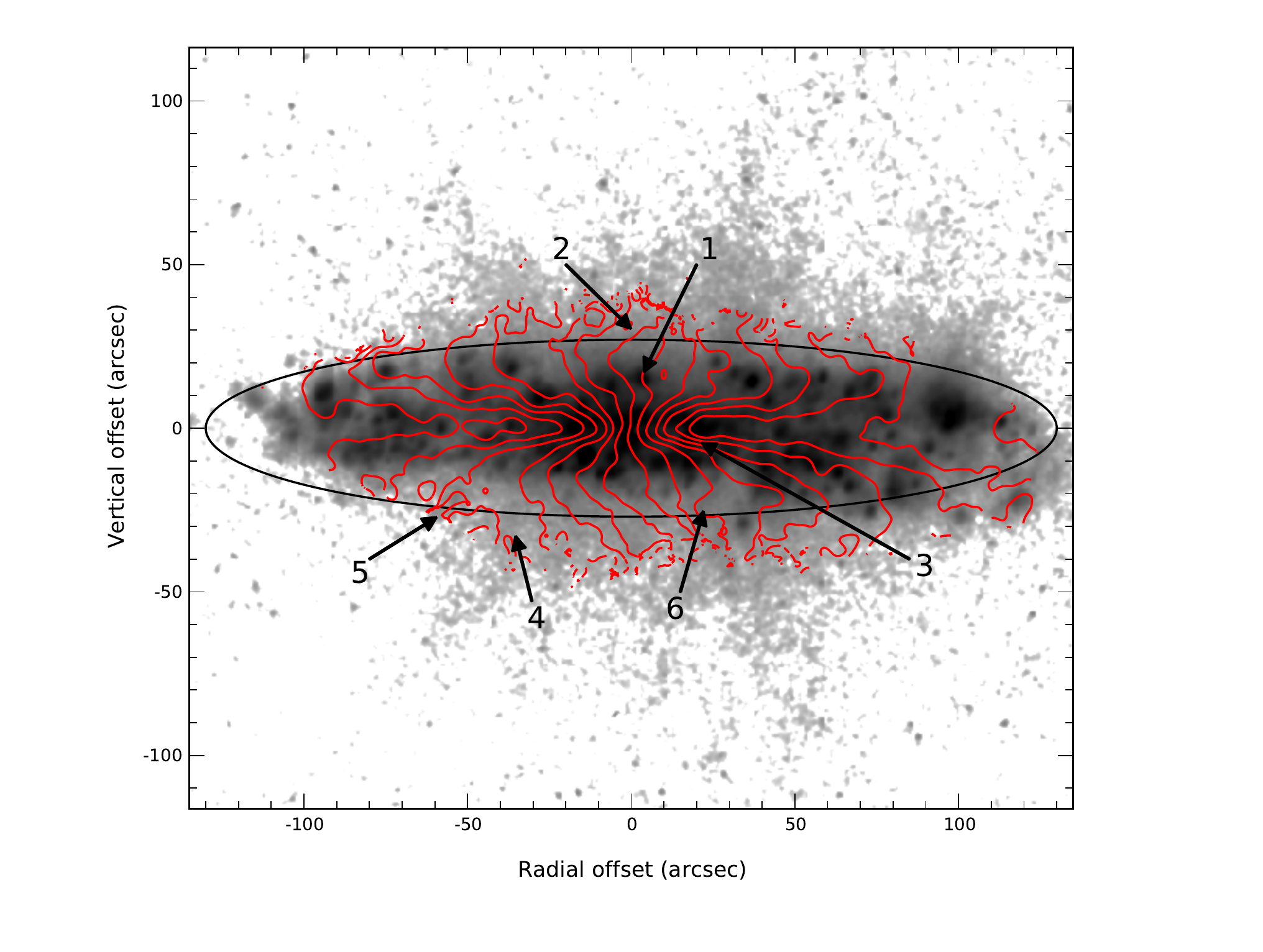}
      \caption{The H$\alpha$ from \citet{dahlem1997} overlaid with the velocity contours of our data in steps of 25\,km s$^{-1}$ around the systemic velocity of 1506\,km s$^{-1}$ and a rough approximation of the disk size by tilting a circle with a diameter of 130\arcsec~to the inclination of the galaxy (78$^\circ$).}
         \label{dahlemhalpha_vel_contours}
   \end{figure*}

To describe the halo features seen in Fig. \ref{ngc4666dispersion} in more detail, a deeper view of the emission in the galaxy is necessary. For this we overlaid the H$\alpha$ image from \citet{dahlem1997} with our velocity image in contours in Fig. \ref{dahlemhalpha_vel_contours}. To guide the eye, the black line representing a circle tilted at 78$^\circ$ with a diameter of 130\arcsec~is drawn as a rough approximation of the disk size.

First of all, we note that because of the lopsidedness of the galaxy the approximation of the disk with a simple ellipse fails on the left side. Furthermore, the above-mentioned influence of the \ion{H}{ii} regions in the velocity field is visible at several points, but more interesting are the structures below -25\arcsec~and above +25\arcsec~vertical offset. In the H$\alpha$ image one can see extraplanar filaments reaching deep into the halo. In comparison to our velocity field one can see that our measurements go into these halo structures but are not as deep as the image by \citet{dahlem1997}. Nevertheless, a small protrusion in the velocity field at a radial and vertical offset of 30\arcsec~also corresponds to an outgoing filament. The reasons that our observations do not show such structures in the halo are on the one hand the cut-off due to the ghost images which would influence the structure in the outer halo and, on the other hand the necessity to have a significant signal above the noise to realize a Gaussian fitting. 

The structure of the galaxy in the lower-left quadrant is extremely interesting. In all other quadrants the ellipse describes the shape of the galactic disk fairly well, but not there. Between the ellipse and the real disk is a distance of about 10\arcsec~which corresponds to 0.9\,kpc. A possible explanation is the lopsidedness and, therefore, also a smaller size of the galactic height of the left part of the galaxy. However, in the linewidth map (Fig. \ref{ngc4666dispersion}) this region shows a strong increase of the linewidth that cannot be explained with our models. Furthermore, in Fig.~\ref{dahlemhalpha_vel_contours} several filamentary structures and knots of higher emission like in the hook-like structure close to position 5 are visible. Therefore, these structures could represent outflowing extraplanar gas.

The behavior of the velocity along the minor axis, especially the double Gaussian structure at region 2, can be understood better when taking into account the filamentary structures above +25\arcsec~vertical offset that can be seen in the H$\alpha$ image. We interpret this gas as part of an outflow. In region 6 no \ion{H}{ii} region is visible, but strong emission of DIG up to a vertical offset of -50\arcsec. Therefore, an overlap of the normal rotation velocity and the velocity of the outflowing gas can be expected there.

\section{Comparison with observations in other wavelengths}

For the final interpretation a comparison with observations in other wavelengths is necessary.

\citet{dahlem1997, dahlem1998} detected extraplanar soft X-ray emission and radio synchrotron emission from the halo and interpreted these as further evidences for a galactic outflow.
\citet{walter2004} detected \ion{H}{i} tidal streams connecting to NGC~4668. Furthermore, they assumed that the high star formation rate is triggered by gravitational interaction of NGC~4666 with the other galaxies in the group. For the \ion{H}{i} \citet{walter2004} determined an inclination for the inner 2\arcmin~of about 70$^\circ$ whereas our determination of the inclination amounts to 78$^\circ$. This results in a velocity offset of about 7\,km\,s$^{-1}$. Taking this into account the maximal rotation velocity in \citet{walter2004} is 173\,km\,s$^{-1}$. This is in good agreement to the value of the inner arcminute in our analysis of the ionized hydrogen and still in agreement to our maximal rotation velocity of 195\,km\,s$^{-1}$ when taking into account the errors.

Although optical observations, including our own observation, do not show any evidence of a strong interaction with other galaxies, such as a tidal streams, NGC~4666 cannot be seen as an isolated galaxy. Especially the lopsidedness of NGC~4666 could be a hint for such a gravitational interaction with other galaxies. It is unlikely that the regular thick gaseous disk found in this study would be caused by environmental effects. But finally one has to take into account also the surrounding galaxies for a complete interpretation. However, the data for such an analysis is currently not available.

\section{Summary and conclusions}

We observed NGC~4666, an edge-on galaxy with a high star formation rate, with a Fabry-P\'erot interferometer to get line of sight velocity information of the ionized hydrogen in the disk and in the halo. Based on these data we calculated the rotation velocities, which result in a flat rotation curve with a rotation velocity of about 195\,km\,s$^{-1}$. 

A comparison of the velocity along the minor axis, which should be constant, with the results of \citet{dahlem1997} yields a good agreement within the inner $\pm$20\arcsec~of vertical offset. The differences at higher vertical offsets can be attributed to additional velocity components, which are present in our data and not in the data of \citet{dahlem1997} as a result of our better spectral resolution. Especially for a positive vertical offset of more than 20\arcsec, a second velocity component with a velocity offset of about -100\,km\,s$^{-1}$ is confirmed, which we interpret as a strong outflow.

To interpret the velocity and the linewidth of the fitted profiles in an appropriate way we created models with different parameters like inclination and scale height, based on the determined rotation velocity. As result of comparing these different models with the data, a combination of two models, one with a thin disk (z~=~0.2\,kpc) and one with a thick disk (z~=~0.8\,kpc), gave the best fit. A first important result is that the double Gaussian x-shaped structure in the center of the linewidth map as well as the double Gaussian structure in the outer regions at negative radial offsets are results of projection effects. The exact shape and the linewidth differs on small scales between data and model, but because of the large number of \ion{H}{ii} regions embedded in the diffuse ionized gas of the disk this is not surprising. We conclude as the main result of this paper the existence of a thin and a thick disk of diffuse ionized gas in NGC~4666, similar to the one in NGC~891 \citep{dettmar1990, rand1990} or in our own Galaxy \citep{Reynolds1990}. Finally, we can conclude that modeling the linewidth map offers a proper possibility of detecting a thin and a thick disk of ionized gas in highly inclined, but not edge-on, galaxies.

Furthermore, we found extraplanar structures, e.g., at a radial offset of 25\arcsec~in both vertical directions, consistent with previous H$\alpha$ observations. Clearly, the star formation in NGC~4666 maintains a diffuse ionized thick disk and drives also a large scale outflow. If there is really a galactic superwind, cannot be concluded with our data. It is worth noting here that the outflow appears driven by a high star forming disk and not a nuclear starburst. This makes NGC~4666 more similar to z\,=\,2 turbulent disks \citep[e.g.,][]{genzel2011} than merger driven nuclear starbursts \citep[e.g.,][]{heckman1990}.

\begin{acknowledgements}
This research has made use of the NED, which is operated by the Jet Propulsion Laboratory, California Institute of Technology, under contract with NASA. 
We acknowledge the use of the HyperLeda data base (http://leda.univ-lyon1.fr). We thank J. L. Gach, P. Balard, and O. Boissin for preparing the instrument for the observations and for repairing a failure in the cooling system during the run. P. Kamphuis thanks the Alexander von Humboldt Foundation for financial support during his stay in Germany. Research in this area at Ruhr-Universit\"at Bochum is supported by FOR 1048. Last but not least, we want to thank the referee for useful comments.
\end{acknowledgements}

\bibliographystyle{aa} % style aa.bst
\bibliography{aa} % your references Yourfile.bib

\end{document}